
\input phyzzx

\def\winf{$w_{\infty}$}
\def\Winf{$W_{\infty}$}
\def\g{{\cal G}}
\def\del{\partial}
\def\tr{{\rm tr}}
\def\nl{\hfill\break}
\def\ni{\noindent}
\def\tA{{\tilde A}}
\def\ta{{\tilde a}}
\def\e{\epsilon}

\hfill{IASSNS-HEP-91/89}\break
\indent \hfill{TIFR-TH-91/61}\break
\indent \hfill March, 1992 \break
\date{}
\titlepage
\title{\bf Classical Fermi Fluid and Geometric Action for $c=1$}
\author{Avinash Dhar$^*$, Gautam Mandal\foot{e-mail: adhar@tifrvax.bitnet,
mandal@tifrvax.bitnet.}}
\address{ Tata Institute of Fundamental Research, Homi Bhabha Road, Bombay
400 005, India}
\andauthor{Spenta R. Wadia
\foot{Supported by DOE grant DE-FG02-90ER40542.}
\foot{e-mail: wadia@tifrvax.bitnet}}
\address{School of Natural Sciences, Institute for Advanced Study,
Princeton, NJ 08540, U.S.A}
\andaddress{ Tata Institute of Fundamental Research, Homi Bhabha Road, Bombay
400 005, India}\foot{Permanent address.}
\abstract{We formulate the $c=1$ matrix model as a quantum fluid and
discuss its classical limit in detail,
emphasizing the $\hbar$ corrections. We view the fermi fluid profiles as
elements of \winf-coadjoint orbit and write down a geometric action for
the classical phase space. In the specific representation of fluid
profiles as `strings' the action is written in a four-dimensional form
in terms of gauge fields built out of the
embedding of the `string' in the phase plane. We show that the
collective field
action can be derived from the above  action provided one restricts
to quadratic fluid profiles and ignores the dynamics of their `turning
points'.}
\endpage

\section{\bf Introduction:}

Two-dimensional string theory provides us with a good laboratory to
discuss principles of string theory and gravity. One of its attractive
features is the high degree of solvability both at the classical and
quantum level. In particular for a flat background metric and a linear
dilaton the theory is equivalent to $c=1$ matter coupled to
two-dimensional gravity
\REF\DNW{S. Das, S. Naik and S.R. Wadia, Mod. Phys. Lett. A4 (1989) 1033.}
\REF\DJNW{A. Dhar, T. Jayaraman, K.S. Narain and
S.R. Wadia, Mod. Phys. Lett. A5 (1990) 863.}
\REF\DDW{S. Das, A. Dhar and S.R. Wadia, Mod. Phys. Lett. A5 (1990) 799.}
\REF\POLA{J. Polchinski, Nucl. Phys. B234 (1989) 123.}
\REF\BL{T. Banks and J. Lykken, Nucl. Phys. B331 (1990) 173.}[\DNW-\BL]
and has a non-perturbative formulation in terms of
the double-scaled $c=1$ matrix model
\REF\BKZ{E. Brezin, V.A. Kazakov and Al.B. Zamolodchikov, Nucl. Phys. B338
(1990) 673.}
\REF\GROSSMILK{D.J. Gross and N. Miljkovic, Nucl. Phys. B238 (1990) 217.}
\REF\PARISI{G. Parisi, Europhys. Lett. 11 (1990) 595.}
\REF\GINZINN{P. Ginsparg and J. Zinn-Justin, Phys. Lett. 240B (1990) 333.}
[\BKZ-\GINZINN].
It is known for a long time that
this matrix model is equivalent to a gas of non-interacting fermions
moving in an external potential\REF\BIPZ{E. Brezin, C. Itzykson, G. Parisi
and J.B. Zuber, Comm. Math. Phys. 59 (1978) 35.}[\BIPZ].
The fermions allow us not only to solve
the theory
\REF\SW{A.M. Sengupta and S.R. Wadia, Int. J. Mod. Phys. A6 (1991) 1961;
A.M. Sengupta, G. Mandal and S.R. Wadia, Mod. Phys. Lett. A6
(1991) 1465.}
\REF\GK{D.J. Gross and I. Klebanov, Nucl. Phys. B352 (1990) 671.}
\REF\MOORE{G. Moore, Rutgers preprint, RU-91-12 (1991).}
\REF\POLB{J. Polchinski, Nucl. Phys. B346 (1990) 253.}[\SW-\POLB]
but also to find a large symmetry group (\Winf)
\REF\DDMW {S. Das, A. Dhar, G. Mandal and S.R. Wadia, ETH, IAS and Tata
preprint, ETH-TH-91/30, IASSNS-HEP-91/52 and TIFR-TH-91/44 (Sept. 1991),
to appear in Int. J. Mod. Phys. A.}
\REF\DDMWB{S. Das, A. Dhar, G. Mandal and S.R. Wadia, Mod. Phys. Lett. A7
(1992) 71.}
\REF\DDMWC{S. Das, A. Dhar, G. Mandal and S.R. Wadia, IAS and Tata
preprint, IASSNS-HEP-91/79 and TIFR-TH-91/57 (December 1991), to appear in
Mod. Phys. Lett. A.}
\REF\MS{G. Moore and N. Seiberg, Rutgers and Yale preprint, RU-91-29 and
Yale-P19-91 (1991).}
\REF\MINIC{D. Minic, J. Polchinski and Z. Yang, Texas preprint, UTTG-16-91
(1991).}
[\DDMW-\MINIC]
that has the interpretation of performing generalized gauge transformation
on the fermions (generalized because it involves both multiplying the
fermion field by a phase and transporting it in space). This symmetry has
also been found in two-dimensional string theory \REF\WITTEN{E. Witten,
IAS preprint IASSSNS-HEP-91/51 (1991).}
\REF\POLKLEB{I. Klebanov and A.M. Polyakov, Princeton University preprint.}
[\WITTEN, \POLKLEB],
and in the collective field theory which approximates the fermion
field  theory \REF\AJ{J. Avan and A. Jevicki, Brown preprint
Brown-HET-801 (1991).} [\AJ]. In particular in [\WITTEN] there is
a realization of vertex operators in terms of a ring of functions
in a two-dimensional phase space and associated vector fields.
See also \REF\ZWEIB{B. Zweibach and E. Witten, IAS preprint
IASSNS-HEP-92/4.} [\ZWEIB]. In
[\DDMW] and [\DDMWC] we explored the consequences of the \Winf\ symmetry
in the quantum theory. It turns out [\DDMW] that the ``classical limit''
of the above symmetry can be understood very naturally in terms of the
classical limit of the fermi theory. In the semi-classical approximation
the states of the fermi theory are represented by a ``fermi fluid'' of
uniform density occupying a certain two-dimensional region (not
necessarily connected or simply connected) in the single-particle phase
plane. Excitations of this state correspond to deformations of the fermi
fluid. Such a concept is well-known in condensed matter physics
\REF\NOZ{See, for example, P. Nozieres, Theory of Interacting
Fermi Systems Pt.2, W.A. Benjamin, INC. (1964), pp 229.} [\NOZ]
and has been applied to the problem at hand in a limited sense in
[\POLB]. In the quantum theory the excitations must satisfy the
condition that the total number of fermions is conserved-- in the
semiclassical approximation this means that the total area of the fermi
fluid must be conserved. Hence the excitations of the quantum theory are
semiclassically represented by area-preserving deformations of the fermi
fluid. If we concentrate for the moment on those deformations that are
differentiable, then we see that the basic excitations of the theory are
area-preserving diffeomorphisms, hence they are elements of \winf. We thus
recover the
classical limit of \Winf. In [\DDMW] we used this observation to represent
fluid profiles by \winf\ `angles' and discussed the dynamics in terms of the
latter.

In this paper we develop in more detail the classical limit of the one
dimensional fermi gas and its correspondence with the quantum theory.
We summarize the results below.

(1) Correspondence between classical and quantum theory: We show that the
$W(p,q,t)$ operator introduced in [\DDMW-\DDMWC] have the interpretation
that the expectation value of its fourier transform in any state $|F>$
gives the quantum distribution function $u_F(x,p)$, satisfying the
property that
$$<F| \int dx \Psi^\dagger(x,t) \hat f(x,-i\del_x) \Psi(x,t)|F>
=\int dx\, dp\, f(x,p) u_F(x,p) \eqn\azero$$
We show that the equation of motion of $W(p,q,t)$ is simply a statement of
Liouville's theorem for the quantum distribution function. We discuss the
classical limit of $u_F(x,p)$ which leads to density functions that are
one inside a region $R$ and zero outside.
We  derive the change of $u_F(x,p)$ as the state $|F>$ is changed to
$U|F>$ ($U$ belonging to the \Winf-group). This gives a natural
representation of \Winf\ in the space of $u_F$'s. We show that the
classical limit of this representation corresponds to a representation of
\winf\ on the classical distribution functions, which coincides with the
representation we had introduced in [\DDMW] and will use below. We also
present the operator algebra of the quantum fluid characterized by
$W(p,q,t)$ and $u(x,p,t)$.

(2) In [\DDMW] we identified the classical phase space of the $c=1$ matrix
model with the space of shapes of an incompressible fermi fluid (fluid
`profiles') in the two-dimensional phase plane of a single fermion. As
suggested by correspondence with the quantum theory, we
represent the fluid profiles by density functions that are characterisitic
functions. The \winf-action on these fluid profiles corresponds to
thinking of the area-preserving diffeomorphism as a hamiltonian flow and
evolving each fermion constituting the fluid under that flow.
We show that there is a natural notion of scalar product
between fluid profiles and canonical transformations
which enables us to identify the
\winf-action on the fermi fluid as a co-adjoint action. This allows us to
write down a geometric action on the classical phase space a la Kirillov.
We find that this action is the classical limit of the action we wrote down
in [\DDMWB] involving elements
of the \Winf-group.

(3) We show that there
is a rather interesting `string representation' of the above
classical dynamics. One parametrizes the boundary of the fermi fluid in
the form of a (closed or open) string. Under this parametrization the
dynamics of the fluid profile resembles a `string theory' in 2 space
dimensions with a built-in reparametrization invariance that follows from
the indistinguishability of fermions.
In order to write down the geometric action mentioned in (2) in
these variables, one needs to introduce a time $t$ and an additional
variable $s$, so that the basic variables of the theory are $x^i(\sigma,
\tau, t,s), i=1,2; x^1=x, x^2=p$ which, at any given $t,s$, describes
two-dimensional region (we consider connected simply connected regions)
occupied by the fermi
fluid which is thought of as the image of a map ($x^i$) from a
two-dimensional parameter space $\sigma, \tau$.  The most interesting
feature of the string representation is that the geometric action is given
by
$$ S_0= \int_M dt\, ds\, d\sigma\, d\tau F_{\sigma\tau}F_{st}
\eqn\zero $$
where $F_{\mu\nu}$ is the field strength of an abelian gauge potential
$A_\mu= \epsilon_{ik} x^i \del_\mu x^k, \;i,k=1,2;\;\mu=1,2,3,4$. We are
using the notation $\xi^\mu= (\sigma, \tau, t,s)$ for coordinates of the
four-dimensional space $M$ (the topology of $M$ is discussed in section 5
below).

We also show that under a
certain `gauge choice' which is valid only in some restrictive class of
configurations (namely that the fluid profile is quadratic, that is, the
fluid boundary is given by $0= F(x,p)$ where $F$ is at most quadratic in
$p$  and that
the turning point of the fluid boundary on the $x$-axis
is static), we
recover the collective field theory action \REF\JS{A. Jevicki and
B. Sakita, Nucl. Phys. B165 (1980) 511.}\REF\DJ{S.R. Das and A.
Jevicki, Mod. Phys. Lett. A5 (1990) 1639.} [\JS, \DJ].

The  preceding couple of paragraphs deal with fermi fluids that occupy one
filled region. The situations with multiple filled regions or regions with
holes (multiply connected regions) correspond, in the
string-interpretation, to splitting and joining of strings, and hence to
an interacting string theory. We should mention that by using the action
in terms of the $U$ variable, it is possible in principle to calculate the
action for a ``pants'' diagram, by invoking $U$-variables which are
perhaps singular and which take a one-string configuration to two-string
configurations. This is similar in spirit to the calculation of a
three-string vertex using the Polyakov path integral.

The plan of the paper is as follows. In section 2 we discuss the
correspondence between classical and quantum theory. In section 3 we
review the Kirillov method
briefly to set up the notation. In section 4, we apply this method
to the case of the fermi fluid and derive the geometric action.
In section 5, we
discuss the string representation, derive the action \zero\ and  show
how to recover collective field theory from this action under the
restricted circumstance mentioned above.

\section{\bf Correspondence between the classical and the quantum theory:}

In [\DDMW-\DDMWC] we introduced the fermion bilinear
$\Phi(x,y,t)=\Psi^\dagger(x,t) \Psi(y,t)$ or its relative
$$W(p,q,t) \equiv \int dx \Psi^\dagger(x,t) \hat g(p,q) \Psi(x,t), \qquad
\hat g(p,q) \equiv \exp(ip \hat X+ iq\hat P) \eqn\aone$$
as the basic dynamical variable of the theory\foot{The definition of
$W(p,q,t)$ given here differs from the one in our earlier papers
[\DDMW-\DDMWC] by a factor of two.}. Since all states of the
fermi theory are obtained by multiple application of the $W(p,q,t)$'s to
the ground state (filled fermi sea)\foot{To see this, note that
elementary particle-hole excitations can be written as linear combination
of $W(p,q,t)$'s ({\sl cf.} [\DDMW].},
it is clear that they provide a complete set of
physical observables.

Let us recall the commutation relation and equation of motion of the
$W(p,q,t)$'s [\DDMW]
$$[W(p,q,t), W(p',q',t)]= 2i \sin[{\hbar\over 2}(pq'-qp')]
W(p+p', q+q',t)
\eqn\atwo$$
$$ (\del_t + p\del_q + q\del_p) W(p,q,t) =0 \eqn\athree $$
In writing the last equation we have used a specific hamiltonian
$h(x,-i\del_x) =(-\del_x^2 - x^2)/2$.

We shall see now that $W(p,q,t)$'s have a rather interesting
interpretation in the classical limit. Let us compute the
expectation value in a state $|F>$ of an observable
$$O_f = \int dx\, \Psi^\dagger(x,t) \hat f \Psi(x,t) \eqn\afour$$
where $\hat f(x,-i\del_x)$ is an operator corresponding to a classical
function $f(x,p)$ in the single-particle phase space. We shall fix the
operator ordering in $\hat f$ by defining
$$\hat f\equiv \int d\alpha d\beta \tilde f (\alpha, \beta) \hat
g(\alpha,\beta), \quad \tilde f(\alpha, \beta)\equiv\int {dx\over 2\pi}\,
{dp\over 2\pi}\, f(x,p) \exp(-i\alpha x-i\beta p) \eqn\afive$$
$\hat f$, defined in this fashion is called a Weyl-ordered
operator. To understand \afive, note that it can be obtained from the
classical relation
$$ f(x,p) = \int d\alpha d\beta \tilde f(\alpha,\beta) g(\alpha, \beta;
x,p), \quad g(\alpha,\beta;x,p) \equiv \exp(i\alpha x+ i\beta p)
\eqn\asix$$
by replacing the symbol $x$ by $\hat X$ and $p$ by $\hat P$ on the right
hand side so that $g(\alpha, \beta; x,p)$ becomes $\hat g(\alpha, \beta)$
(recall the definition of $\hat g$ in \aone). As an example, if $f(x,p)=
xp^2$ then $\tilde f(\alpha,\beta)= (-i\del_\alpha)\delta(\alpha)
(-i\del_\beta)^2 \delta(\beta)$, so that $\hat f= \int d\alpha\, d\beta\,
(-i\del_\alpha)\delta(\alpha) (-i\del_\beta)^2 \delta(\beta) \exp(i\alpha
\hat X + i\beta \hat P) = \hat X (\hat P)^2 - i \hat P= (1/3)[\hat X(\hat
P)^2 + \hat P\hat X\hat P+ (\hat P)^2\hat X]$ which is perhaps a more
familiar statement of Weyl ordering.

The expectation value of $O_f$ in a state $|F>$ is then
$$\eqalign{
<F| O_f|F>= &\int d\alpha\, d\beta\,<F|\int dx\, \Psi^\dagger(x,t) \hat
g(\alpha, \beta) \Psi(x,t)|F> \tilde f(\alpha, \beta)\cr
=&\int_{\alpha, \beta} <F|W(\alpha, \beta)|F> \tilde f(\alpha, \beta) \cr
=& \int_{x,p} <F|u(x,p)|F> f(x,p)\cr}
\eqn\aseven$$
where $u(x,p,t)$ is the Fourier transform of $W(\alpha,\beta,t)$:
$$u(x,p,t)=\int_{\alpha,\beta}\, \exp(-i\alpha x- i\beta p)
[\int dy\, \Psi^\dagger(y,t) \hat g(\alpha,\beta)
\Psi(y,t)] \eqn\aeight$$

Equation \aseven\ is rather interesting. It tells us that the expectation
value of the operator $O_f$ can be exactly expressed by a phase
space integral of $f(x,p)$ with a density function
$$u_F(x,p,t)\equiv <F|u(x,p,t)|F> \eqn\anine$$
Since this statement is valid at the quantum level, we call $u_F(x,p)$
the \underbar{quantum} \underbar{distribution} \underbar{function} in the
state $|F>$.

Thus, we find that $W(p,q,t)$ is simply the  Fourier transform of the
quantum distribution function {\sl operator} $u(x,p,t)$
$$ u(x,p,t)= \int d\alpha\, d\beta\, \exp(i\alpha x+
i\beta p) W(\alpha, \beta,t) \eqn\aten$$
whose expectation value gives us the quantum distribution function
$u_F(x,p,t)$.

Note that the equation of motion \athree\ for the $W$-operator implies the
following equation for $u$ (using \aten)
$$(\del_t+ p\del_x+ x\del_p)u(x,p,t)=0 \eqn\aeleven$$
which is simply the statement of Liouville theorem for the quantum
distribution (for the hamiltonian $h(x,p)= (p^2- x^2)/2$).

\ni \underbar{Relation to the ``first-order density matrix'':}\nl

\ni Given a many-body wave-function $|F>= \sum F(x_1, \cdots, x_N)|x_1,
\cdots, x_N>$ the first-order density matrix is defined as
$$\eqalign{
\phi_F (x,y)=&\int dx_2\,dx_3\cdots dx_N F^*(x, x_2,x_3,  \cdots, x_N)
F(y, x_2, x_3, \cdots,x_N)\cr
&=<F|\Phi(x,y)|F>, \quad \Phi(x,y)= \Psi^\dagger(x) \Psi(y)\cr}
\eqn\atwelve$$
The usefulness of this quantity is that a single-particle operator like
$O_f= \sum_i \hat f(x_i, -i\del/\del x_i)= \int dx\,\Psi^\dagger(x,t) \hat
f \Psi(x,t) $ is given by
$$<F| O_f |F> = \sum_{x,y} <x| \hat f|y> \phi_F (x,y) \eqn\athirteen$$
By using \afive\ in \athirteen\ we get
$$<F| O_f|F>= \int_{x',p'} f(x',p') [\int_{x,y} \phi_F (x,y) <x| \tilde
{\hat g}(x',p')|y>]  \eqn\afourteen$$
where $\quad \tilde {\hat g}(x,p) \equiv\int_{\alpha, \beta} \hat
g(\alpha, \beta) \exp(-i\alpha x- i\beta p)$.\nl
\ni Comparing with the previous expression of $<F|O_f|F>$ we find that
the quantum distribution function is a simple transform of the
first-order density matrix:
$$u_F(x,p)\equiv <F|u(x,p)|F> = \int_{x_1, y_1} K(x,p; x_1,
y_1) \phi_F (x_1, y_1) \eqn\afifteen$$
where
$$K(x,p; x_1, y_1)\equiv <x_1| \tilde {\hat g}(x,p)|y_1> =
\exp(-ip(x_1-y_1)) \delta({x_1+ y_1\over 2}-x)
\eqn\asixteen$$
This kernel actually expresses the transformation between the two
different forms of the fermion bilinear, $\Phi(x_1, y_1,t)$ and
$u(x,p,t)$.

\ni \underbar{The semi-classical limit}:\nl

\ni Let us try to calculate $u_F(x,p)$ using \afifteen, when $|F>$ is
the ground state. For simplicity, let us consider a system of free
particles first (potential$=0$). It is easy to calculate $\phi_F (x_1,
y_1)$ using plane waves for the single-particle wavefunctions:
$$\phi_F (x_1, y_1)= \sum_{k<k_F} \exp[ik(x_1-y_1)] \eqn\aseventeen$$
This gives
$$u_F(x,p) = {\rm const.}\; \theta(p_F-p) \eqn\aeighteen$$
The constant ensures that the integral $\int dx\, dp\, u_F(x,p)$ which
measures the total fermion number, is $N$ (in the case of double scaling
where $N\to\infty$ the normalization has to be appropriately redefined).

Equation \aeighteen\ is the first example of recovering a classical result
from the quantum distribution function. We recall that semiclassically the
ground state of a fermi system corresponds to  filling up the region in
the phase space corresponding to $h(x,p)< e_F$ where $h(x,p)$ is the
energy function. In other words, the classical density function is
$$u^C_F(x,p) = {\rm const.}\; \theta(e_F-h(x,p)) \eqn\anineteen$$
where the constant is again determined by the condition that the integral
of $u^C_F$ should reproduce the total fermion number. We can see that for
the free particle case ($h(x,p)= p^2$) \anineteen\ coincides with
\aeighteen. In other words, the quantum distribution function coincides
with the classical distribution function.

In general (for a more general potential and more general states) the
classical distribution function is given by
$$u^C_F(x,p)= \chi_R(x,p) \eqn\atwozero$$
where $\chi_R(x,p)$ is the characteristic function for some region $R$ in
the phase space ($\chi_R(x,p)=1$ if $(x,p)\in R$, $=0$ otherwise).
$R$ describes the region occupied by the fermions in the state $|F>$--- in
general this is only an approximate concept, and hence the quantum
distribution function has an $\hbar$-expansion:
$$u_F(x,p)= \chi_R(x,p) + O(\hbar) \eqn\atwoone$$
corresponding to a quantum softening of the step function. Of course at
finite temperature there is a softening of the step function even at the
classical level.
In the above we have ignored the constant multiplying $\chi_R(x,p)$.

In the case of $h= (p^2- x^2)/2$ the quantum distribution function is
indeed not exactly
equal to the characteristic function, and has non-trivial $\hbar$
corrections as in \atwoone, but we skip the details here.

We see therefore that the natural set of observables in the quantum
theory, namely $u_F(x,p)= <F|u(x, p)|F>$ correspond in the
classical limit to the set  of density functions $\chi_R(x,p)$ for all
regions $R$ in the phase plane. In section four we will define the
classical phase space of the fermi system in terms of characteristic
functions.

Before we end this section, let us show how the \Winf-transformations in
the quantum theory give rise to \winf-transformations on the fluid regions
$R$ in the classical theory modulo $\hbar$-corrections.

We have shown in [\DDMW] that the fermion fock space forms an irreducible
representation of the \Winf-algebra. This is a consequence of the fact
that an arbitrary state in the theory can be reached from the fermi ground
state by particle-hole pair excitations, which in turn can be expressed as
linear combination of the \Winf-generators. This implies that an arbitrary
state $|F>$ in the theory can be written as
$$|F> = U|F_0> \eqn\atwotwo$$
where $U$ is a \Winf-group element:
$$U= \exp[i \int_{p,q} \epsilon(p,q) W(p,q,t)] \eqn\atwothree$$

Let us ask the following question: if a state $|F_0>$ (not necessarily the
ground state) is changed by a
\Winf-transformation according to \atwotwo, how does the corresponding
quantum distribution function change?  By definition
$$u_0\to u_F(x,p)= <F_0| U^\dagger u(x,p)U|F_0>
\eqn\atwofour$$
We have abbreviated $u_{F_0}$ by $u_0$.

Let us consider the case when $U= 1+ iH= 1+i \int_{\alpha,\beta}
\epsilon(\alpha,\beta) W(\alpha, \beta)$ with $H$ infinitesimal.
\atwofour\ now reads
$$\delta u_0(x,p)= -i \int_{\alpha, \beta} \epsilon(\alpha, \beta)
<F_0| [W(\alpha, \beta), u(x,p)] |F_0>\eqn\atwofive$$
We can evaluate the commutator by using the structure constants \athree\
of the \Winf-algebra. We have
$$\eqalign{
[W(\alpha, \beta,t), u(x,p,t)]
& =2i \int_{\alpha',\beta'} \sin[{\hbar\over
2}(\alpha\beta'-\alpha'\beta)]
\exp(-i\alpha' x- i\beta' p) W(\alpha+\alpha', \beta+ \beta') \cr
& = 2i \sin [{i\hbar\over 2}(\alpha \del_p-\beta \del_x)][
e^{i\alpha x+ i\beta p} u(x,p,t)]\cr
}
\eqn\atwosix$$

Putting this in \atwofive\ one gets
$$\delta u_0(x,p) = i\int_{\alpha, \beta} \epsilon (\alpha, \beta)
\delta_{\alpha, \beta} u_0(x,p) \eqn\atwoseven$$
where
$$\delta_{\alpha, \beta} u_0 (x,p) = \int_{x',p'} M_{\alpha, \beta}(x,p;
x',p') u_0(x',p') \eqn\atwoeight$$
with
$$\eqalign{
M_{\alpha, \beta}(x,p; x',p')=& 2i\int_{\alpha',\beta'}
\sin[{\hbar\over 2}(\alpha\beta'-\alpha'\beta)]
\exp[-i\alpha' x- i\beta' p+ i(\alpha+ \alpha') x'+i(\beta+ \beta') p']\cr
=& 2i \sin[{i\hbar\over 2}(\alpha \del_p - \beta \del_x)]
[\delta(x-x')\delta (p-p')\exp(i\alpha x'+ i\beta p')]\cr}
\eqn\atwonine$$
Equations \atwoeight\ and \atwonine\ define the representation of
\Winf-algebra on the quantum distribution functions.

To understand the classical limit of equations \atwoeight\ and \atwonine\
we need to make an $\hbar$-expansion of these equations. The net result is
that the sine function gets expanded in power series, the first term
representing the classical limit. Using this, we get
$$M_{\alpha, \beta}(x,p;x', p') = -\hbar(\alpha\del_p- \beta\del_x) [
\delta'(x-x')\delta(p-p')\exp(i\alpha x'+ i\beta p')] + O(\hbar^2)
\eqn\athirty$$
The last equation implies
$${1\over i\hbar}
\delta_{\alpha, \beta} u_0= \{\exp(i\alpha x+ i\beta p), u_0\}_{PB}
+ O(\hbar) \eqn\athreeone $$
In other words, if one performs a \Winf-transformation on the quantum
distribution function by a generator corresponding to the single-particle
operator $\exp(i\alpha \hat X+ i\beta \hat P)$ the result, to leading
order in $\hbar$, is to send $u_0(x,p) \to u_0(x',p')$ where
the points $x',p'$ are obtained from $x,p$ by a canonical transformation
under the corresponding classical function $\exp(i\alpha x+ i\beta p)$.
The division by $i\hbar$ is related to the difference between commutators
and classical Poisson brackets. Note that the $\{\}_{PB}$ here denotes
Poisson bracket in the single-particle phase space. Equation \athreeone\
is remarkable because (a) it relates the Poisson bracket structure in the
many-body phase space to that in the single-particle phase space and (b)
it sets up a correspondence between \Winf-trnasformations on the left hand
side to \winf-transformations on the right hand side.

Thus we proved that \Winf-transformation on the quantum distribution imply
\winf-transformation on the classical distribution. The coadjoint orbit
construction in section four for the classical case
will precisely use this transformation.

To end this section we mention that a detailed discussion of the operator
algebra of the quantum fluid, considered briefly here, is under
preparation
\REF\DMWB{A. Dhar, G. Mandal and S.R. Wadia, Tata preprint, in
preparation.}[\DMWB]. We summarize the operator algebra in the following:
$$\eqalign{
[W(p,q,t), W(p',q',t)]= & 2i \sin[{\hbar\over 2}(pq'-qp')]
W(p+p',q+q',t)\cr
[W(p,q,t), u(p',q',t)]= & 2i \sin[{i\hbar\over 2}(p\del_{p'}-q\del_{q'})]
[e^{i(pq'+qp')} u(p',q',t)] \cr
[u(p,q,t), u(p',q',t)]= & 2i\sin[{\hbar \over 2}(\del_p \del_{q'}- \del_q
\del_{p'})] [\delta (p-p') \delta(q-q') u(p',q',t)] \cr}
\eqn\athreetwo$$
where $u(p,q,t)$ in the last two lines denotes $u(x,p,t)$ with $x=q$.

\REF\KIRILLOV{See for example,
A.A. Kirillov, {\sl Elements of the Theory of Representations} (1976);
A. Alekseev and S. Shatasvili, Nucl. Phys. B 323 (1989) 719;
A. Alekseev, L. Faddeev and S. Shatasvili, J. Geom. Phys. 1 (1989) 3;
B. Rai and V.G.J. Rogers, Nucl. Phys. B341 (1990) 119; J. Avan and A. Jevicki,
Brown preprint.}
\section{\bf Review of the Kirillov method[\KIRILLOV]:}

Traditionally, the Kirillov method is a method of inventing physical
systems that form representations of a  group $G$. The steps are the
following:

(a) Find a dual $\Gamma$ to the Lie algebra $\g$, that is, invent a linear
space $\Gamma$ so that there is a scalar product $<x|u>$ between elements
$x\in \Gamma$ and $u\in \g$.

(b) Define an action of $G$ (and consequently of $\g$) on $\Gamma$ by the
rule  $G\owns U: x \to x^U\equiv\tA(U).x \in \Gamma$
where $\tA(U).x $ is defined by
$$ < \tA(U).x| u > = < x | U u U^{-1}>\; {\rm for\; all} \, u\in \g
\eqn\one$$
The group action $\tA(U)$ satisfying \one\ is called a co-adjoint action
and the property \one\ is called the co-adjoint property (for obvious
reasons, since the right hand side of \one\ uses the adjoint action on the
Lie algebra $\g$).\nl
\ni Note that $\tA$ has the property $\tA(UV)= \tA(U)\tA(V)$.\nl
\ni We shall also use the infinitesimal version of $\tA$, namely if $U=
\exp (tv)$ then $\ta(v).x\equiv \lim_{t\to 0} (d/dt) \tA(U).x$. In this
limit, the coadjoint property reads as
$$ <\ta(v).x| u>= <x| [v,u]>\; u,v\in G, x\in \Gamma \eqn\two$$
\ni \underbar{Coadjoint orbit}: for any given $x_0\in \Gamma$ the set of
points $\tA(U).x_0$ obtained by applying all group elements $U\in G$ to
$x_0$ is
called the coadjoint orbit of $x_0$ (denoted $C(x_0)$).

(c) Parametrize points of  $C(x_0)$ by group elements, as follows. Suppose
$x\in C(x_0)$. Clearly, by definition of $C(x_0)$, there is at least one
group element $U$ that has brought us from $x_0$ to $x$, i.e.
$$ \tA(U). x_0 = x \eqn\onea$$
Actually there
would usually be an ambiguity in the definition of $U$, because if some
$U$ applied to $x_0$ gives $x$, then so would  $UV$ where
$$\tA(V)x_0= x_0\eqn\oneb$$
The set of such $V$'s forms a subgroup $H\subset G$, called the
stability subgroup of $x_0$. This means that points of $C(x_0)$ are
characterized by an equivalence class $UH$; in other words, $C(x_0)= G/H$.

(d) Let us assume that we have fixed the ambiguity by choosing a
particular element out of each equivalence class. Then each point $x\in
C(x_0)$ has an image $U \in G$. Similarly, curves $x(t)\in C(x_0)$ have
images $U(t) \in G$. It is easy to see that the tangent vector $dx/dt$
corresponds to the tangent vector $dU/dt$. By the well-known isomorphism
between tangent vectors in a group manifold and elements of its Lie
algebra we know that a tangent vector $dU/dt$ at the point $U(t)$
corresponds to the Lie algebra element $(dU/dt)U^{-1}\equiv u_t$.
Note that $dx/dt= \ta(u_t).x$.

Kirillov's prescription for the symplectic form is the following. Suppose
we have two tangent vectors at the point $x\in \Gamma$, given by
$t_1=\ta(u_1). x$ and $t_2=\ta(u_2). x$. Then the syplectic form
$\Omega(x)$ is defined by
$$\Omega(x). (t_1, t_2) \equiv <x| [u_1, u_2]> \eqn\three$$
One can check that $\Omega$ is anitysmmetric and closed.

(e) The way one arrives at the classical action for a path $x(t)\in
\Gamma$ is by extending the path $x(t)$ to a two-dimensional region
$x(t,s)$ whose boundary is $x(t)$. The classical action is simply the
integral of the syplectic form over this two dimensional region\footnote*{
Actually, this corresponds to the $p\dot q$ part of the Lagrangian, we
will include the hamiltonian ($H dt$) part shortly.}. Consider
a little two-dimensional region formed by the two tangent vectors at x:
$$ dx/dt= \ta(\del_t U U^{-1}).x, \; dx/ds= \ta(\del_s U U^{-1}).x
\eqn\four$$
By the above definition for the symplectic form, the action is therefore
$$ S_0 = \int dt ds <x | [\del_t U U^{-1}, \del_s U U^{-1}]>\eqn\five $$

The above action seems to depend on the extension $x(t,s)$ of the original
path in the $s$ direction. The fact that $\Omega$ is closed ensures that
the action does not change under small deformations of the map $x(t,s)$
which keep the boundary $x(t)$ invariant.

One can formally integrate the Lagrangian with respect to $s$ (in
the absence of any topological obstruction). Thus
$$\eqalign{S_0 =& \int dt ds
<x_0 | [U^{-1} \del_t U,U^{-1} \del_s U ]> \cr
 = & \int dt <x_0 | U^{-1} \del_t U> \cr
 = & \int dt <x| \del_t U U^{-1}> \cr
 }\eqn\six $$
where in writing the first and the third lines line one has used the
co-adjoint property and the relation between $x(t,s) $ and $x_0$.

The question of well-definedness of the action: note that we have managed
to write an action \five\ or \six\ in terms of the $U$-variable. However,
the action must have an invariance with respect to the change $U\to UV$
where $V$ satisfies \oneb, because both $U$ and $UV$ correspond to the
same point $x$ in the configuration space, and the action should depend
only on the path drawn in the configuration space (and not in $G$). In
other words, in case of \five, we demand that
$$ S_0[U(s,t)] = S_0[U(s,t) V(s,t)] \, {\rm where}\, \tA(V(s,t)).x_0=x_0
\eqn\sixa$$
and in case of \six\ we demand that
$$S_0[U(t)] = S_0[U(t)V(t)] \, {\rm where} \, \tA(V(t)).x_0= x_0
\eqn\sixb$$
We call these criteria the criteria for well-definedness of the action.
Equation \sixa\ is easily satisfied by using the coadjoint property of the
scalar product. \sixb\ is more tricky \REF\Bala{A.P. Balachandran,
G. Marmo, B.S. Skagerstam and A. Stern, {\sl Gauge Symmetries and Fibre
Bundles}, Springer Lecture Notes in Physics, Vol. 188.}
[\Bala] and instead of the Lagrangian remaining invariant it changes by a
total derivative in time.

\underbar {Adding  the Hamiltonian piece}:

So far we dealt with only the `symplectic form' or the $p\dot q$ part
of the action. In general, we will have a hamiltonian piece, where the
hamiltonian $h$ corresponds to  a given element of the Lie algebra, $h\in
\g$. We want to add $-dt H$ term in the action so that we get an equation
of motion
$${ dx\over dt} = \ta(h). x \eqn\sixc$$
The additional piece is
$$S_h= -\int dt <x| h>\eqn\sixca$$
and the total action
$$S= S_0 + S_h \eqn\sixcb$$
It is easy to check that the equation of motion following from the above
action is indeed \sixc.

This is the standard story.

Our approach will be slightly different, mainly in emphasis. We will
consider the  physical system, say $Q$, as already given to us, with a
specified action of the relevant group $G$. We will try to see if we can
invent a notion of scalar product between elements of $Q$ (to be suitably
identified as a subset of some linear space $\Gamma$) so that under that
scalar product the group-action satisfies the co-adjoint property. Then we
can use the Kirillov device to write down a classical action on $Q$ (or
more specifically on coadjoint orbits in $Q$). In the case of the fermi
fluid, $Q$ is going to be the space of fluid profiles (= the space of
characteristic functions),  and the cojoint orbits $\subset Q$ going to be
fermi
fluids of a given area. We shall see that there is an obvious scalar
product between $Q$ and the elements of \winf\ algebra, indeed if we use
the identification of the space of fluid profiles  with the space of
characteristic functions then $Q$ is naturally imbedded in the space
$\Gamma$ of distributions and the above scalar product is then inherited
from $\Gamma$.

\underbar{A Toy Example: single classical spin as co-adjoint orbit of
$SU(2)$}

Let us explain a toy example. This is the problem of a single classical
spin, characterised by a ``spin" vector $x^i\in R^3$ satisfying the the
constraint of unit norm $\vec x. \vec x=1$. The configuration space of the
spin is $Q= S^2$. How does one write a natural action for this spin?

First remark that there is an obvious action of $SU(2)$ on this spin,
namely rotation of the spin. To be precise, if we denote an element of
$su(2)$ (the algebra) as $u= u_i \sigma^i$ where $\sigma$ are the Pauli
matrices, then
$$\ta(u). \vec x= \vec u\times \vec x \eqn\seven$$

One way of quickely arriving at this equation is to
think of $R^3$ as identified with the Lie algebra $su(2)$ where $\vec x$
is identified with a matrix $X= x^i \sigma_i$. In this case, the rotation
by $u$ of $X$ is given by $\ta(u). X= [u,X]$ using the natural action of
Lie algebra on itself. \seven\ is obtained
by using the fact that $[u,X] = \vec
u \times \vec x. \vec \sigma$.

The above identification  of $R^3$ with $su(2)$ also suggests a natural
scalar product between elements of $R^3$ (and hence of $Q\subset R^3$) and
the elements of $su(2)$, namely
$$<x | u> = {\rm tr}(X u) = \vec x. \vec u \eqn\eight$$

Clearly the above action \seven\ of $su(2)$ satisfies the co-adjoint
property under the scalar product \eight. The co-adjoint orbits are
spheres in $R^3$ of constant radii. By choice, our config. space $Q$ is
the coadjoint orbit characterised by unit radius.

Classical action: Let us try to construct the classical action for a
periodic path $\vec x(t), \vec x(0)=\vec x(1)$
which in our case is a closed path drawn
on the sphere $Q$. The classical action is the integral of the symplectic
form over the ``filled circle"\footnote*{There is an ambiguity here
regarding whether we want to fill the `inside' or the `outside' of the
circle, but we'll see that the symplectic form is an integral form which
means that the results will differ only by an integal multiple of $2\pi$,
hence in computing $\exp(iS)$ we would not see any difference}.
We denote the
filled circle by $\vec x(t,s)$, where $s\in [0,1]$, $s=1$ describing the
boundary.

Consider now the infinitesimal region formed by the two tangent vectors
$d\vec x/dt$ and $d\vec x/ds$ drawn at the point $\vec x(t,s)$.
The integral of the
symplectic form over this infinitesimal region is
$$ ds \, dt \, <x| [u_t, u_s]> \eqn\nine$$
where $u_t, u_s$ are elements of the $su(2)$ algebra defined by the
equations
$$\eqalign{ \ta(u_t). \vec x =& \vec u_t\times \vec x=
{d\vec x \over dt} \cr
\ta(u_s). x =& \vec u_s\times \vec x=
{d\vec x \over ds} \cr} \eqn\ten$$

It is easy to solve \ten, giving
$$\eqalign{
\vec u_t=& \vec x\times {d \vec x\over dt} \cr
\vec u_s=& \vec x\times {d \vec x\over ds} \cr}
\eqn\eleven$$
We have used the notation $u_t= \vec u_t.\vec \sigma$, etc. and the fact
that the tangent vectors $d\vec x/dt$ and $d\vec x/ds$ are perpendicular
to $x$ (also that $\vec x.\vec x=1$).

Using these one arrives at a classical action
$$S_0= \int dt\, ds\,  \vec x. \vec {dx\over dt} \times {dx\over ds}
\eqn\twelve $$
which is the famous solid angle action for a classical spin.

A typical hamiltonian term corresponds to inclusion of a magnetic field
which may be viewed as either a vector $\vec B$ or equivalently as a
matrix $B= \vec B.\vec\sigma$. The corresponding term in the action is
$$S_h= \int dt <x| B> = \int dt \vec x.\vec B \eqn\thirteen$$

\section{\bf The fermi fluid as co-adjoint orbit of \winf:}

We consider the space $Q$ of fermi fluid profiles as the classical phase
space. In [\DDMW] we derived the transformation of the fluid profile under
area-preserving diffomorphisms (equivalently, canonical transformations).
In section two we re-derived this result by looking at the classical limit
of the quantum distribution function.

Following the discussion in section 2, we shall parametrize the fermi
fluid as follows. If the fluid occupies a
region $R$ in the phase plane we shall characterize it by a density
function $\chi_R(x,p)$ defined by

$$\eqalign{\chi_R(p,x)=& 1, {\rm if}\,
(x,p) \in R \cr =& 0, \;{\rm otherwise}\cr} \eqn\fourteen$$

Thus, our classical phase
space is the space of density functions
$Q= \{\chi_R, R\subset {\bf{\rm R}}^2\}$ where
this $R$ can be any two dimensional subset of the phase plane.

The group we are concerned with is the group of area-preserving
diffeomorphisms. The Lie algebra consists of hamiltonian flows under
arbitrary functions of the phase plane. Thus we shall parametrize the Lie
algebra elements $u$ by functions $f(x,p)$ or vector fields
$X_f = \del_p f \del_x - \del_x f \del_p$.  When we think of the
functions as Lie algebra elements, we define the Lie bracket as identical
to the Poisson bracket. Note that
$$[X_f, X_g] = X_{\{f,g\}_{PB}}$$.

The action on fluid elements is specified by considering the action on the
corresponding density functions:
$$ \ta(f).\chi_R(x,p)= - (\del_p f \del_x - \del_x f
\del_p)\chi_R(x,p)\equiv -X_f\chi_R(x)\eqn\fifteen$$
Let us justify the above definition. Note that for infintesimal $f$, the
above can be re-stated as
$$\chi_R(x,p) + \ta(f)\chi_R(x,p)= \chi_R(x',p')\eqn\sixteen$$
where $(x',p')$ is the point obtained by evolving $(x,p)$ for unit time
under the hamiltonian $-f$: $x'= x- \del_p f, p'= p +\del_x
f$. Now instead of evolving the  cooridnates $x,p$ we might alternatively
evolve the region $R$. We know that if we evolve both $R$ to $R'$ and the
coordinate $x,p$ to $x',p'$ then $\chi_R(x,p)=\chi_{R'}(x',p')$. This also
shows that $\chi_R(x',p')= \chi_{R'}(x,p)$ where $R'$ is now obtained
by  evolving $R$ under the hamiltonian $f$ for unit time. Thus we get the
result that, for infinitesimal $f$,
$$\chi_R(x,p) + \ta(f)\chi_R(x,p)= \chi_{R'}(x,p)\eqn\seventeen$$
In other words, under the \winf\ transformation $f$, the region $R$ evolves
to a new region $R'$ as if each fermion inside evolves under the
hamiltonian $f$. In the next sections we shall use  \seventeen\ instead of
\fifteen\ to find the action of infinitesimal \winf\ transformations in the
different parametrizations.

Note that the above representation of \winf\ is presicely the one
obtained in section two as a limiting case of \Winf-transformation on the
quantum distribution function.

The action of finite group elements, like $U(t)= \exp(-iX_f t )$ is given
by $$\tA(U).\chi_R(x,p)= \exp(it (\del_p f \del_x - \del_x f \del_p))
\chi_R(x,p)\eqn\eighteen$$

\underbar{The scalar product:}

We define now a notion of a scalar product between the elements $X_f$ of
\winf\ and fluid elements $\chi_R(x,p)$. A natural scalar product is:
$$<\chi_R(x,p)| X_f> = \int_{{\bf{\rm R}}^2}
\chi_R(x,p) f(x,p)=\int_R f(x,p) \eqn\twenty$$
In other words the scalar product counts the total amount of $f(x,p)$
contained in the fermi fluid.

Note that if we think of elements of \winf\ as simply functions $f(x,p)$
on the phase plane, the obvious dual $\Gamma$ is the space of `generalized
functions' or distributions (indeed distributions are defined that way).
The scalar product of a distribution $D$ with a function $f$ is by
definition the integral of the distribution with $f$ as the `test'
function. Our density function $\chi_R$ is defined as the characteristic
function corresponding to the region $R$, hence the space of fluid
profiles $Q$ (defined as the space of density functions) is naturally
embedded in the linear space $\Gamma$ of distributions. Indeed one can
extend the definition of \winf\ action on $\Gamma$ in the obvious way. To
be precise, the action of $X_f$ on a distribution $D$ would be given by
$$<\ta(f).D | g>= <D| X_f.g>\equiv <D| \{f,g\}_{PB}>\eqn\twoone$$

\twoone\ makes it clear that the above scalar product satisfies the
co-adjoint property. The coadjoint orbits are fluids of the same area.
{}From the point of view of the fermion theory, this is a natural
consequence of fermion number conservation.

\underbar{Classical action:}

We want to compute the classical action for a fluid
trajectory $R(t)$ (which we assume periodic for the moment), given by
density functions $\chi_{R(t)}$. We assume that $R(t)$ has the same area
for all $t$, so that it is lying on a single coadjoint orbit. We ``fill
in'' the one-dimensional trajectory to a two-dimensional one as mentioned
in the last section: the fluid profiles are called $R(s,t)$ and the
corresponding density functions $\chi_{R(s,t)}$. The action is given by
the following two dimensional integral
$$S_0= \int dt\, ds\, <\chi_{R(s,t)}| [\del_t U U^{-1}, \del_s U U^{-1}]>
\eqn\twoonea$$

The group element $U(s,t)$ is defined to be the one which brings us from a
certain ``base fluid profile'' $R_0$ to the current one $R(s,t)$:
$$\tA(U(s,t)).\chi_{R_0}= \chi_{R(s,t)} \eqn\twooneaa$$
As explained earlier, $U(s,t)$ as defined by \twooneaa\ is ambiguous upto
right multiplication by elements $V(s,t)$ which do not move $R_0$:
$$\tA(V(s,t)). \chi_{R_0} = \chi_{R_0} \eqn\twooneab$$
This means that our configuration space (space  of fluid profiles of a
given area that form an orbit of \winf) is actually a coset $G/H$ where
$H$ is the set of $V$'s satisfying \twooneab. One can check that the action
written above satisfies the creiterion of well-definedness expressed in
\sixa.

As mentioned in the general outline in the previous section, we can
transpose the $U$-action to rewrite \twoonea\ in terms of $\chi_{R_0}$.
The result, after one partial integration with respect to $s$ (in
the absence of a topological obstruction), is
$$\eqalign{
S_0= &\int dt <\chi_{R_0}| U^{-1} \del_t U > \cr
=& \int <\chi_R| \del_t U U^{-1}> \cr}\eqn\twooneb$$

The hamiltonian piece:  if the fluid profile is evolved by a hamiltonian
function $h(x,p)$ in the single-particle phase space (for instance, $h=
(p^2- x^2)/2$), then \sixca\ becomes
$$ S_h= -\int dt <\chi_R| X_h> \eqn\twoonec$$
By using the coadjoint property of the scalar product we can rewrite this
as
$$S_h= -\int dt <\chi_{R_0}| U^{-1} X_h U > \eqn\twooned$$
The right way to interpret the above expression is: first think of $U$ as
made of exponential of differential operators (like $U=\exp(X_f)$ etc.) so
that $U^{-1} X_h U$ is also a differential operator, of the form $X_g$ for
some $g$. Then $<\chi_{R_0} | U^{-1} X_h U>$ is actually defined to be
$<\chi_{R_0} | g>$.

The total action is given by
$$S=S_0+S_h= \int dt\, <\chi_{R_0} | U^{-1} \del_t U+ U^{-1} X_h U>
\eqn\twoonedzero$$

\underbar{Correspondence with the action in [\DDMWB]:}

Let us make a brief remark to make correspondence with the \Winf-action
that we wrote down in [\DDMWB]. The latter action was
$$\eqalign{
S=S_0 + S_h, \, S_0=& \int dt\, \tr(\Lambda U^{-1} \del_t U),\cr
S_h= &\int dt\, \tr(\Lambda U^{-1} \bar A U) \cr}
\eqn\twoonedc$$
where $U=\exp(i \int \epsilon(p,q) \hat g(p,q))$ is an element of
\Winf-group, viewed as an operator in the single-particle Hilbert space
or equivalently as an infinite dimensional matrix
$U_{xy}= <x| U|y>$. $\Lambda$ is a fixed matrix, and $\bar A$ is simply
the hamiltonian operator $h$ appearing in the fermion theory.

It is clear that the above action is the Kirillov action for the
co-adjoint orbit of $\Lambda$ under the group \Winf. The coadjoint action
is defined by identifying the dual of the algebra with itself (the scalar
product being defined as the {\sl trace}).

We are not going to prove the equivalence between \twoonedc\ and
\twoonedzero\ in great detail here; we indicate the steps instead.
Basically,
we found in section two that the scalar product
$<\chi_R| X_f>$ is merely the classical limit of the quantity
$<u_F| X_f>\equiv\int_{x,p} f(x,p) u_F(x,p)$ where $u_F(x,p)$ is the
quantum distribution function in the state $|F>$ (the state $|F>$ here
is {\it
defined} by the property that $u_F(x,p) = \chi_R(x,p) + O(\hbar)$). Now
the latter quantity is expressible as
$\int_{x,y}f_{xy} \phi_F (x,y)$ where $f_{xy}= <x| \hat f|y>$ (see
equation \athirteen).
Now consider the hamiltonian term $<\chi_{R_0}|U^{-1} h U> = <\chi_R|h>$
in \twoonedzero. By the above remarks, this is the classical limit of
$\tr (h \phi_F )$ where we have suppressed the symbols $x,y$,
treating them as
matrix indices. If we now use the fact that an arbitray state $|F>$ state
can be written as $|F>= U|F_0>$, then it is easy to deduce that
$\phi_F = U^{-1} \phi_{F_0} U$. Thus, if we identify $\Lambda(x,y)=
\phi_{F_0} (x,y)$, then the hamiltonian piece in \twoonedc\ equals that
in \twoonedzero\ plus order ($\hbar$) terms. Similar remarks hold for the
kinetic term, though the proof is a little more lengthy\footnote*{Note
that we have used the same notation $U$ above for the single-particle
operator\nl
\ni $U=\exp[i\int dp\, dq\, \epsilon(p,q) \hat g(p,q)]$ and for the
many-body operator $U=\exp[i\int dp\, dq\, \epsilon(p,q) W(p,q)]$.}.

\section{\bf The String Representation:}

In this section we shall employ a different, in a sense more intrinsic,
representation of the fermi fluid, which we shall call the ``string
representation". Let us consider for the moment a fermi fluid which
occupies one single filled region (that is, a connected simply connected
two-dimensional region of the phase plane). One example is the ground
state distribution. If the hamiltonian is $h= (p^2+ x^2)/2$
then the ground
state distribution is a fermi fluid filling the region $ p^2+ x^2 \leq
2\mu$ where $\mu$ is the fermi energy. If the hamiltonian is
$h=(p^2-x^2)/2$
(and we restrict to $x<0$, cf. [\DDMW]) then the ground state distribution
is $x\leq -\sqrt{p^2-2\mu}$. We point out that the choice of signature
(Euclidean or Minkowski) in the 2-dimensional target space dictated in
[\DDMW] the choice of the hamiltonian (actually, in the Euclidian case
the hamiltonian was $h= - (p^2 + x^2)/2$
so that the fermi fluid filled the
outside of the circle mentioned above-- this difference will not matter
for most part of our discussion below and we'll choose to ignore it).
As is clear from the example of the ground state distributions, the
boundary of the fermi fluid can either be closed or open.

In the ``string representation" we describe the fermi fluid by specifying
the boundary as the map $x^i(\sigma)$ from a one-dimensional parameter
space $\sigma\in [0,2\pi]$ or $\sigma\in (-\infty, +\infty)$ (depending on
whether the boundary is closed or open respectively) to the phase plane
$x^i,\, (x^1=x, x^2=p)$. Indeed  since we are considering filled (simply
connected) regions, let us invent a two-dimensional parameter space
$\sigma, \tau$ (which is a disc or a half-plane) which maps onto the
two-dimensional region filled by the fermi fluid. The map is
$(\sigma,\tau) \mapsto x^i(\sigma,\tau)$.
If we parametrize the disc by $\tau\in [0,1]$ then
the boundary is given by $\tau=\tau_0\equiv 1$ (in the case of the
half-plane, parametrized by $\sigma\in (-\infty,\infty), \tau\in
(-\infty,0]$ the boundary is $\tau=\tau_0\equiv 0))$. The `string' is simply
the image of the one-dimensional boundary $\tau=\tau_0$ of the parameter
space.

Action of \winf\ on the `string':

Again, according to \seventeen, we evolve the point $(x,p)$ of the phase
plane (``target space") under the hamiltonian $h=f$:
$x^i \to x^i+\epsilon^{ik} \del_k f$, which means that entire two-dimensional
image $x^i(\sigma, \tau)$ changes by
$$\ta(f). x^i(\sigma,\tau) \equiv \epsilon^{ik}\del_k f (x(\sigma,\tau),
p(\sigma,\tau)) \eqn\twofour$$
and the one-dimensional image of the $\tau$-boundary (the string
$x^i(\sigma, \tau_0)$) transforms by
$$\ta(f). x^i(\sigma,\tau_0) \equiv \epsilon^{ik}\del_k f (x(\sigma,\tau_0),
p(\sigma,\tau_0)) \eqn\twofive$$

We should remark that it is only the `string' that is the real dynamical
variable. If we make any transformation that moves the fermions {\it
inside} the fermi fluid, leaving the boundary unchanged, then the physical
state of the system is unchanged (because the phase space density is
unchanged-- the original reason for this of course is the
indistinguishability of identical fermions). In concrete terms, any
transformation in the target space $(x,p)\to (x',p')$ that leaves the
image $x^i(\sigma,\tau_0)$ invariant, does not change the ``fermi fluid".
Indeed the statement is even stronger. Even if the map $x^i(\sigma,
\tau_0)$ is changed in a manner such that a reparametrization of the
boundary $\sigma\to \sigma'$ can account for the change, then we havent
really changed the fluid profile.

It is interesting to ask what are the canonical transformations that
change the string only upto reparametrization. The answer is, all those
transformations $h$ that satisfy
$$\ta(h). x^i(\sigma,\tau_0)\propto \del_\sigma x^i(\sigma, \tau_0)
\eqn\twosix$$
The proportionality `constant' can be a function of $\sigma$ (indeed if the
function is $g(\sigma)$ the reparametrization that is implied here is
$d\sigma'/d\sigma= g(\sigma)$). To see what these functions precisely are,
let's combine \twofive\ and \twosix. We get
$$\del_i h \del_\sigma x^i= \del_\sigma h (x^i(\sigma, \tau_0))=0
\eqn\twoseven$$
which implies that the ``string'' is a  surface of constant $h$. It is clear
that for the fermi fluid in the ground state, any function of the energy is a
candidate $h$. Since such $h$'s do not change the fermi fluid, the space
of fermi fluids is actually a coset $G/H$ [\DDMW].

The scalar product: in this parametrization we have
$$<\chi_R| f> = \int_R dx\, dp\, f(x,p)= \int_D d\sigma d\tau
\e_{ik}\del_\sigma x^i \del_\tau x^k f(x(\sigma,\tau), p(\sigma, \tau))
\eqn\twoeight$$
where $D$ denotes the entire parameter space $\sigma, \tau$ (disc in case
of closed strings and half-plane in case of open strings).

To write the action consider $x^k$ as a function of time $t$ and
the additional
variable $s$. Then, \four\ applied to our case is
$${\del x^k\over \del b}=\ta(f_b).x^k=\epsilon^{kl}\del_l f_b,
\;b=s,t \eqn\twonine$$

The classical action \twoonea\ now reads
$$S_0= \int dt\, ds\int_D d\sigma\, d\tau\, (\del_\sigma f_s \del_\tau f_t
- (s\leftrightarrow t)) \eqn\twoten$$
Using \twonine, we can evaluate the partial derivatives involved in
\twoten. The result is
$$S_0= \int dt \, ds\, d\sigma\, d\tau\, F_{\sigma\tau}F_{st}
\eqn\twoeleven$$
 The field strength is defined as
$$F_{\mu\nu}= \e_{ik} \del_\mu x^i \del_\nu x^k\eqn\twotwelve$$
which can be derived from a gauge potential
$$A_\mu = \e_{ik} x^i \del_\mu x^k \eqn\twothirteen $$
Here $\mu,\nu$ run over all coordinates of the four-dimensional space
$\xi^\mu=(s,t, \sigma, \tau)$.

Some interesting features of this gauge theory are:

(1) The gauge transformations $A_\mu\to A_\mu+ \del_\mu \theta$ correspond
to making canonical transformations $x^i\to x^i+ \e^{ik}\del_k f$ where
$$\theta= (2- x^i\del_i) f \eqn\twofourteen$$

(2) In the notation of differential forms, the gauge potential is given by
a one-form
$$A= A_\mu d\xi^\mu= \e_{ik} x^i dx^k=xdp-pdx \eqn\twofifteen$$
and the gauge field by the two-form
$$F= \e_{ik} dx^i \wedge dx^k=dx(\xi)\wedge dp(\xi)
=X_*(\Omega) \eqn\twosixteen$$
where $X_*(\Omega)$ denotes the pull-back of the symplectic form $\Omega$
in the phase plane onto the four-dimensional space $\xi^\mu$ (the map $X$
refers to the embedding $\xi^\mu \to x^i(\xi)$). In other words, the
symplectic form $\Omega= dx\wedge dp$ in the target space induces a
symplectic form in the four-dimensional space. Our field strength is
precisely equal to that.

In this notation the action looks like
$$S_0= \int_M d\sigma\, d\tau\, ds\, dt\, (X_*(\Omega))_{\sigma\tau}
(X_*(\Omega))_{st}  \eqn\twoseventeen$$
Note that the  induced symplectic structure
degenerates where the embedding map $X$ is singular. This is precisely
what happens at the turning points of the profiles. We shall come back to
the issue of the turning point shortly.

The hamiltonian term \twoonec\ in the string representation can be written
as
$$ S_h=\int_M B\wedge X_*(\Omega) \eqn\twoeighteen$$
where $B$ is defined as the two-form $B= (\del_x h \del_s x + \del_p h
\del_s p) ds\wedge dt$.

Let us now see under what conditions we can derive the collective field
theory from the above considerations. For this purpose it is more useful
to use the form \twooneb\ of the classical
action and use the string parametrization in it. The second line of
\twooneb\ reads as
$$S_0= \int dt \int_R f_t  \eqn\threeone $$
where $f_t$ satisfies
$$ \del_i f_t = \e_{ik} \del_t x^k  \eqn\threetwo $$

Let us convert the area integral over the region $R$ in \threeone\ into a
line integral by thinking of $f_t$ as a magnetic field and inventing a
vector potential $a_i, i=x,p$. That is, let us find $a_i$ such that
$$\del_x a_p - \del_p a_x = f_t \eqn\threethree$$
In that case \threeone\ becomes
$$ S_0 = \int dt \int_{\del R} (a_p dp + a_x dx) = \int dt d\sigma (a_p
\del_\sigma p+ a_x \del_\sigma x) \eqn\threefour$$

Now this reduction is true for any choice of $a_i$ which satisfies
\threethree. Let us choose the gauge $a_p=0$, then from \threethree\
we have
$$a_x= -(\del_p)^{-1} f_t  \eqn\threefive$$
and so we get
$$S_0=\int dt \, d\sigma \del_\sigma x [ -(\del_p)^{-1} f_t]
\eqn\threesix$$
In order to make connection with the collective field action
we would now like to specify points on the
``string" by their $x$-values rather than by the parameter $\sigma$. In other
words, we attempt a change of variable in \threesix\ from $\sigma$ to
$x(\sigma)$. Now since $x(\sigma)$  is actually $x(\sigma,t)$, how does
one make the change of variable at all $t$?

We will see that the way out is suggested by reparametrization invariance.
Note that a given fluid profile (more precisely the density function
$\chi_R$) does not change if one simply makes a reparametrization of the
boundary of the fluid. Therefore given a classical path of a string
described as $x^i(\sigma, t)$, physically it is the same as another path
$x^i(\sigma'(\sigma,t), t)$. Indeed one can check that the above action
\threesix\ is invariant under $x^i(\sigma, t) \to
x^i(\sigma'(\sigma,t),t)$.  Note that we are talking here about arbitray
{\sl time-dependent} reparametrizations  $\sigma \to \sigma'(\sigma,t)$.

Now though in the initial parametrization $\sigma$, $x(\sigma, t)$
depended on $t$, by changing over to $\sigma'$ we may try to keep
$x(\sigma'(\sigma, t), t)$ invariant in time by
compensating between the two sources of time-variation. In other words,
let us see if we can satisfy
$$ {dx\over dt}
= \del_t x + \del_t \sigma' \del_\sigma x =0 \eqn\threesixa$$
Clearly except  when $\del_\sigma x=0$, we can choose the
reparametrization $\sigma'(\sigma,t)$ to satisfy \threesixa. Which means
that except at these points we have
$$x(\sigma'(\sigma,t), t) = x(\sigma) \eqn\threeeight $$

What is the significance of the points where $\del_\sigma x$ vanishes?
Well, these are precisely the turning points of the fluid profile on the
$x$-axis. The ``gauge  choice'' \threeeight\ cannot be validly made at
these turning points. Physically this means that at all points except
where the fluid boundary has turning points in the $x$-direction, one can
always make a combination of vertical and horizontal motions (of the
fermions living at the boundary) appear as a purely
vertical motion by giving the fermion a suitable velocity component along
the boundary (such motions do not change the state of the system, hence
one is always allowed to add them without changing anything). At the
turning point, since the tangent is vertical, adding any amount of
tangential motion will change only the vertical component, and will never
``gauge away'' a horizontal component of the velocity there.

Thus, \threeeight\ is a valid gauge choice only when one restricts to
fluid profiles  which never move their $x$-turning points, that is, the
`turning points' are allowed to move purely vertically (mathemtically, one
is saying that for \threesixa\ to be valid at a point where $\del_\sigma
x=0$ one must have $\del_t x=0$).
Let us assume such a restriction for the moment so that \threeeight\ is
valid. Let us also
assume that in the entire range of $\sigma$ there is only
one turning point on the $x$-axis, that is, there is only one value
$\sigma= \sigma_0$ such that $\del_\sigma x=0 $ at $\sigma_0$\foot{We are
considering for the moment the case of open string, i.e. non-compact
$\sigma$; for the closed-string case we have to assume that there are two
turning points}. This assumption is equivalent to the assumption of
quadratic profiles[\DDMW], that is, that the fluid boundary is given by an
equation $F(x,p)=0$ where $F$ is at most quadratic in $p$.
Now, in the intervals
$(-\infty, \sigma_0)$ and $(\sigma_0, \infty)$ the map $\sigma\to
x(\sigma)$ is separately invertible. We shall call the inverse maps
$\sigma_-(x)$ and $\sigma_+(x)$ in the two intervals, respectively.  We
shall also use the notation $p(\sigma_+(x),t)= p_+(x,t)$ and
$p(\sigma_-(x),t)=p_-(x,t)$.

Note that using \threetwo\ and \threeeight\
we have $\del_p f_t= \dot x=0 \Rightarrow
(\del_p)^{-1} f_t= p f_t$. Since $\del_x f_t= - \dot p$, we have
$f_t= -(\del_x)^{-1} \dot p$. In these relations, $p(\sigma, t)$ is to be
interpreted as $p_\pm(x,t)$ depending on whether $\sigma\ge\sigma_0$ or
$\sigma\le \sigma_0$.

Putting in all of the above, we see that \threesix\ reduces to
the "kinectic" term of the collective field theory action:
$$ \int dt \int dx \, [p_+ {1\over \del_x}{dp_+\over dt}-
p_- {1\over \del_x}{dp_-\over dt}]$$
The hamiltonian term also agrees using similar reasoning and in
this way we get the complete collective field theory action:
$$S=\int dt \int dx \, [p_+ {1\over \del_x}({dp_+\over dt})-
p_- {1\over \del_x}({dp_-\over dt}) + {p_+^3 - p_-^3\over 6}- {x^2\over 2}
(p_+-p_-)]$$

The above method of derivation clearly indicates the limitation of the
collective field description of the theory. Besides the unwarranted
restriction to quadratic profiles one needs to assume that the
$x$-turning point remains static. In general the fluid profile will
move in such a way that the turning
point itself will be dynamical--- the
collective field description clearly misses
this dynamics of the turning point.

\section{\bf Concluding remarks:}

We hope that our classical action \twoonea (or \twoseventeen)
brings new insights into formulating
general principles of classical two-dimensional string theory.
We  wish to emphasize that the ``classical'' action for the quantum
\Winf-symmetry and the corresponding action for its classical limit \winf\
are different. In fact, the ``classical'' action corresponding to \winf\
is only the leading term (in powers of $\hbar$) of the ``classical''
action  corresponding to \Winf. Hence in this circumstance one will not
obtain the correct quantum theory by quantizing this classical action.
Such a circumstance also occurs in string theory. See, for example
\REF\ERIK{E. Verlinde, IAS preprint IASSNS-HEP-92/5 and references
therein.}[\ERIK].

\ni{\bf Acknowledgements:} We would like to thank K.S.Narain, N. Nitsure,
B.Rai and T.R. Ramadas for discussions. We wish to thank S.P. deAlwis for
pointing out an error in an earlier version of the manuscript.
The work of SRW was supported in part by
DOE grant DE-FG02-90ER40542. AD and SRW would like to thank the
International Centre for Theoretical Physics for hospitality
during the completion of this work.

\ni Note added: While completing this work we received
\REF\IKS{S. Iso, D. Karabali and B. Sakita, City College
preprint, 1992.} [\IKS] which
also discusses the fluid picture and the limitations of the collective
field approach.
\refout
\end